\documentclass[aps,prd,twocolumn,groupedaddress,nofootinbib]{revtex4-1}
\usepackage{graphicx}
\usepackage{dcolumn}
\usepackage{bm}
\usepackage{braket}
  
\begin{document}
\preprint{APS/123-QED}

\title{No Firewalls or Information Problem for Black Holes Entangled with Large Systems}

\author{Henry Stoltenberg and Andreas Albrecht}
\affiliation{%
University of California at Davis; Department of Physics
\\One Shields Avenue; Davis, CA 95616
}%

\begin{abstract}
We discuss how under certain conditions the black hole information
puzzle and the (related) arguments that firewalls are a typical feature of black
holes can break down. We first review the arguments of AMPS favoring firewalls, focusing on
entanglements in a simple toy model for a black hole and the Hawking
radiation. By introducing a large and inaccessible system entangled
with the black hole
(representing perhaps a de Sitter stretched horizon or inaccessible part of a landscape) we show
complementarity can be restored and
firewalls can be avoided throughout the black hole's evolution.  Under
these conditions black holes do not have an ``information problem''.
We point out flaws in some of our earlier arguments that such entanglement
might be generically present in some cosmological scenarios, and
call out certain ways our picture may still be realized. 

\end{abstract}

\maketitle
\section{Introduction and Background} 
The prediction that black holes emit Hawking Radiation led to the understanding that black
holes weren't eternal and would eventually evaporate~\cite{Hawking:1974sw}. This presented
an  apparent contradiction between quantum mechanics and general
relativity with respect to the information stored within the black
hole~\cite{Hawking:1976ra, Page:1979tc}. The so-called ``information paradox'' concerns the information
encoded in the initial quantum state of whatever collapsed to form the
black hole (as well as anything that passed through the black hole's
horizon). From the no-hair theorem, the properties of the black hole
are independent of the details of the collapse. This would suggest
that the Hawking Radiation which depends on properties of the black
hole horizon would carry with it no information from the initial
quantum state. Once the black hole has evaporated away completely the
information would be irreversibly lost which would suggest non-unitary
evolution for black holes. 

An apparent solution to restore unitarity was to present corrections to the horizon 
allowing the radiation to carry information away. In a coarse grained description, 
information would seem to disappear but in a more fine grained description the correlations 
between the radiation in the end state would still exist. This solution is realized in 
Black Hole Complementarity~\cite{Susskind:1993if} which is constructed such that no observer 
 can see any contradictions. The postulates of Black Hole Complementarity are:
\\
\\ 
Postulate 1:
The process of formation and evaporation of a black hole, as viewed by
a distant observer, can be described entirely within the context of
standard quantum theory. In particular, there exists a unitary
S-matrix which describes the evolution from in-falling matter to
outgoing Hawking-like radiation. 
\\ 
\\
Postulate 2:
Outside the stretched horizon of a massive black hole, physics can be described to good approximation by a set of semi-classical field equations.
\\
\\
Postulate 3:
To a distant observer, a black hole appears to be a quantum system with discrete energy levels. The dimension of the subspace of states describing a black hole of mass, $M$ is the exponential of the Bekenstein entropy $S(M)$. 
\\ 
\\
In~\cite{Almheiri:2012rt} it was pointed out that there is an implicit additional postulate which is a realization of the equivalence principle from classical GR:
\\
\\
 Postulate 4:
``No drama'' at the horizon. 
\\ 
\\
In other words, from the perspective of an observer freely falling into a black hole, the horizon would not be a special location. Spacetime would appear to be locally flat and for a massive enough black hole, the tidal forces would not be especially large.

A key concept in complementarity is that it is not possible to write
down a quantum state that simultaneously describes the interior and
the exterior of the black hole. From the viewpoint of complementarity,
this would be considered unphysical since those regions are causally
disconnected. Different frames such as ones that just describe the
exterior (an outside observer's frame) and a frame that describes the
interior (a freely falling frame) would have complementary
descriptions. For the outside observer there exists an object, the
quantum mechanical stretched horizon which stands in for the classical
event horizon.  Classical notions of spacetime are expected to break
down at the stretched horizon. The
stretched horizon exists just outside where the classical event
horizon would have been and from
the perspective of external observers is a real physical object with
a temperature, coarse grained thermodynamic entropy and quantum
state. From the perspective of the outside observer, the stretched
horizon thermalizes with in-falling matter and eventually radiates
away the information. From the perspective of a freely falling
observer, there is no stretched horizon. Instead, the horizon is just
like any other location in space and the matter 
simply falls into the black hole's interior. 

Complementarity seems to solve the information problem by preventing a
simultaneous description of the interior and exterior. An observer
falling into the black hole would observe a seemingly contradictory
description to what the observer outside observes. However it has been
argued since the two can never communicate, no issues arise from allowing
both to observe different but self-consistent physical
descriptions.  

In a paper by Almheiri, Marolf, Polchinski and Sully (AMPS) ~\cite{Almheiri:2012rt},
a thought experiment showed that the four postulates of Black Hole
Complementarity seemed to be inconsistent.
 By considering a potential
experiment that could be done by an observer collecting radiation
throughout the history of the black hole's evaporation and then
falling through the horizon, they have discovered a frame where
``quantum monogamy'' (a basic technical feature of quantum mechanics)
would seem to be violated. They considered the least radical
modification would be to do away with ``no-drama'' and propose a
firewall, a barrier of high energy quanta which in-falling matter and
observers would encounter.\footnote{
	Braunstein comes to similar conclusions, reasoning that
        vanishing entanglement across the horizon results in high
        energy particles, an ``energetic
        curtain''~\cite{Braunstein:2009my}.}\footnote{
	Marolf and Polchinski have presented firewalls as typical features of black holes in dual field theories independent of the quantum monogamy arguments ~\cite{Marolf:2013dba}. We do not address these arguments in this paper and focus on the original AMPS thought experiment.}  

In our discussion, we consider a black hole that is entangled with a
large {\em inaccessible} system. The system could be a de 
Sitter stretched horizon, as proposed for example in de Sitter Equilibrium (dSE)
cosmology~\cite{Albrecht:2004ke,Albrecht:2009vr,Albrecht:2011yg,Albrecht:2013lh}
which use as our main illustration.  Equally well, the entangled
system could be a string landscape or multiverse as long as
it exhibits the key properties of being very large and inaccessible
to the observer.\footnote{In some pictures of the string theory
  landscape the inaccessibility property is not realized, for example~\cite{Susskind:2007pv,Freivogel:2006xu}.} 

In a de Sitter equilibrium universe, the observed universe evolves from a
fluctuation in the Gibbons-Hawking radiation of de Sitter space. One
could equally well simply consider a single black hole that forms in
this manner.  In a
picture where the de Sitter horizon is a quantum stretched
horizon, one would expect the state of the fluctuation to
exhibit entanglement with it. Unlike entanglements that have
previously been considered~\cite{Maldacena:2013xja,Susskind:2014rva,Page:2013dx}, the de 
Sitter horizon is a system that could not be measured. 
Specifically, we do not expect there is any way for an
observer to find the information in the de Sitter
stretched horizon which is entangled with a black hole and then jump into the
hole.  This feature allows us to evoke complementarity to avoid the
possibility of observing  quantum inconsistencies.
The size of the de Sitter horizon
system also dwarfs anything in the interior de Sitter space, allowing
there to always be
plenty of states to provide entanglement. We will show explicitly how
these features help in a toy model. The large dimension of the de
Sitter stretched horizon space is part of the reason we expect the
entanglement information to be inaccessible (arguing along similar lines
to~\cite{Parikh:2008iu}).

The question of whether our analysis is applicable in realistic
cosmologies is a tricky one which we discuss in Section~\ref{CaD}.

\section{Toy Model for An Evaporating Black Hole}
\label{sec:bh}
We consider a very simple toy model with a small number of states
that we can write out explicitly.  The full toy model space ``$U$'' can be
written as 
\begin{equation}
U = A \otimes B \otimes C \otimes D \otimes E \otimes F
\label{Udef}
\end{equation}
where each of the subspaces is a two state system. Taking the exterior
viewpoint, these subspaces will either represent degrees of freedom 
of the stretched horizon or ``particles'' of emitted Hawking
radiation. These roles will change as the hole evaporates, and we will
often lump together the subspaces that make up the stretched horizon
under the label ``$H$''. 

First we'll use this toy model to review the conventional
information problem for just a black hole, not in the presence of a de
Sitter stretched horizon. 
Assuming the entire evaporation process is unitary, if the black hole
begins in a pure state it will evolve into a pure state of
radiation. Our toy model system will end with only the six (one bit) ``particles''
in the final state.   
By unitarity, the initial black hole state need not be
described by a larger Hilbert space. Thus our toy model has dimension
$2^6$. 

Generically chosen pure states are known to have entanglement properties
of subsystems related to subsystem sizes.  For subdivisions into a large and small
subsystem, the small subsystem will tend to have little entanglement
within itself (that is, entanglement among its own subsystems). In
contrast, the large subsystem will have a lot of entanglement among
{\em its} subsystems.  The small subsystem will tend to be
``maximally entangled'' with the large one, but not vice versa. In the
Appendix we explain some of these points using simple illustrations. 
Page~\cite{Page:1993df} developed these insights and used them to argue the following:
Consider the time when the initially pure black hole has
radiated roughly half its entropy (often referred to as the Page
time). Not long after the Page time the already emitted radiation will
represent a larger subspace than the black hole.  Thus, most of the
entanglement for the black hole will be with the 
previously emitted radiation (referred to as the ``early radiation''). Also,
there will be very little entanglement between subsystems of
the black hole. Upon completion of the black hole decay, this translates to there being very little
entanglement between subsystems of the ``late radiation''. These arguments
are a result of the assumption that once radiation has gotten
sufficiently far from the black hole, it no longer interacts with the
black hole system and therefore entanglements cannot change.  The toy
model is not complex enough to give a detailed account of the
evolution that exhibits these entanglement properties. Instead we
simply write down toy model states ``by hand'' that reflect
appropriate entanglements for different stages of the evolution. 

We now consider the time (after the Page time) when the black hole has emitted four
particles (represented by subsystems $F$, $B$, $E$ and $A$, given in
order of emission\footnote{The particles are not
  emitted in alphabetical order so as to conform with other
  conventions used for this topic.}). These four
particles are the early radiation.
We expect the remaining black hole subsystem ($H$ =
$C \otimes D$) to be highly entangled with the early radiation.  A
maximally entangled state between the two can be expressed in terms of
entanglement between just two of the early radiation particles and $H$,
and we use subsystems $E$ and $F$ for this purpose. The same general
arguments from Page lead us to expect the remaining
early radiation particles to be maximally entangled with
each other.  While the Page arguments require a larger space to
operate generically, we enforce this feature explicitly on the toy
model by choosing a maximally entangled state for the remaining early
radiation particles $A$ and $B$.

The toy model state
\begin{eqnarray}
(\frac{1}{\sqrt{2}}\ket{\uparrow_A\downarrow_B}-\frac{1}{\sqrt{2}}\ket{\downarrow_A\uparrow_B})\otimes
\ket{\Psi_{H,E,F}}
\label{BHevapState}
\end{eqnarray}
has all the properties listed above to represent a black hole in late
stages of decay.  In particular, particles $A$ and $B$ appear in a
Bell state which exhibits maximal
entanglement. Maximal entanglement between the stretched horizon in
space $H$, and the system of $E \otimes F$  requires a pure state
describing $H$, $E$ and $F$ combined. This state is written as
$\ket{\Psi_{H,E,F}}$. 

Looking at the time when particle $A$ is just leaving the stretched
horizon, we can consider an observer that has collected all the
radiation up to this point who then falls into the black hole, collecting
particle $A$ and attempting to view what lies beyond the horizon. For
this observer there is no stretched horizon and instead, according to
complementarity she finds in the
interior of the black hole a re-expression of the wave function
that was used to describe the horizon (contained within
$\ket{\Psi_{H,E,F}}$). 

A vacuum state should be a pure state. The
purity of the low energy vacuum we expect the falling observer to
experience in the region of 
the horizon implies 
that localized subsystems where one is inside and the other outside the horizon will typically
be entangled.  To the infalling observer particle $A$ represents modes
just outside the horizon. 
To encounter a vacuum, this observer would need to see those modes 
entangled with modes just inside the horizon (call them $C$, which is
part of $H$). This is
where the problem arises. We have already enforced that $A$ be maximally
entangled with $B$. Those particles could have been measured or brought
into a black hole by this
observer. Complementarity requires
local unitarity which would not
allow us to change the entanglement between $A$ and $B$. However, the form
of Eqn.~\ref{BHevapState} (established to meet other requirements as
detailed above) explicitly does not allow {\em any} entanglement between $A$ and
$H$.  This feature is an aspect
of quantum monogamy. With $C$ part of $H$, there is thus also no entanglement between $A$ and $C$. 
Since we have just argued that such an entanglement is required of a
vacuum state, complimentarity appears to
be in conflict with the no drama we expect for the infalling
observer. 

Complementarity allows for quantum monogamy to be broken as long as it
is not violated within a single causal patch. A quantum system could
be maximally entangled with multiple systems as long as no observer could
ever encounter this contradiction. The AMPS thought experiment
provided an example where to enforce ``no drama'' quantum monogamy
would have to be violated within a single causal patch (seen by the falling
observer). Once maximal entanglement between $A$ and $B$ has been established,
nothing else can be entangled with $A$. Since systems $A$, $B$ and $C$ can all be
encountered by a single observer this flexibility afforded by
complementarity does not help.

\section{Toy Model for An Evaporating Black Hole in dS}
\label{sec:bhds}
We consider a black hole (far from the Nariai limit\footnote{The
	Nariai limit places the largest possible black hole in de 
	Sitter space such that the black hole's horizon area approaches the
        area of the de Sitter horizon. We will consider black holes much smaller than this limit,
	giving horizons that are clearly separated and vastly different in
	size.  Very close to the Nariai limit our arguments may break
        down.}) formed by a fluctuation in the Gibbons-Hawking radiation in de Sitter space. We think of the de Sitter
space as having a stretched horizon much as we did for the black
hole.  We expect the Gibbons-Hawking radiation to be strongly entangled
with the stretched horizon. We note that the
entropy associated with the de Sitter horizon is much greater than that of
the Gibbons-Hawking radiation or black hole we are considering within the
space, and thus the stretched de Sitter Horizon subspace will 
have dimension many orders of magnitude greater
than that of the black hole or any other system. 
Consider a black hole forming from Gibbons-Hawking radiation 
which begins maximally entangled with the de Sitter horizon. An
observer outside of and in the rest frame of the hole will eventually observe a time after the
black hole has decayed but before the decay products
reach the de Sitter stretched horizon. It might seem that
the decay products would have similar properties to
Hawking radiation discussed in the previous section.  But in fact unitarity
will ensure that entanglement with the de Sitter stretched horizon will
persist through black hole formation and decay, and become a feature
of the decay products of the black hole once the decay is complete.  In this case we
expect the final state to be a mixed state of the radiation entangled
with the de Sitter stretched horizon. The Hawking radiation produced by the
decaying black hole will always be maximally mixed. 

We can consider similar thought experiments as before.  Here 
the Hawking radiation is always maximally mixed, so the
focus on times after the Page time (which we used in the previous
section) will not be needed in this discussion. With this simplification
we only need to consider {\em two} radiated particles ($A$ and $B$), and
again we chose $B$ to be emitted earlier than $A$. Since they are part of
a maximally mixed system $A$ and $B$ will not have any
entanglement with each other and the combined system of $A$ and $B$ will 
be in a mixed state. Thus the density matrix that describes $A$ and
$B$ combined will be diagonal with equal valued entries on the
diagonal. All this is a consequence of basic facts about
quantum states (connected to the topic of monogamy) which are reviewed
in the Appendix.  A completely general way to write the state of the
entire system including the de Sitter horizon, black hole horizon and
radiation (which all together are in a pure state) is 
\begin{eqnarray}
&\frac{1}{2}\ket{1_H}\otimes(\frac{1}{\sqrt{2}}\ket{\uparrow_A\downarrow_B}-\frac{1}{\sqrt{2}}\ket{\downarrow_A\uparrow_B}) \nonumber\\
+&\frac{1}{2}\ket{2_H}\otimes(\frac{1}{\sqrt{2}}\ket{\uparrow_A\downarrow_B}+\frac{1}{\sqrt{2}}\ket{\downarrow_A\uparrow_B}) \nonumber\\
+&\frac{1}{2}\ket{3_H}\otimes(\frac{1}{\sqrt{2}}\ket{\uparrow_A\uparrow_B}-\frac{1}{\sqrt{2}}\ket{\downarrow_A\downarrow_B}) \nonumber\\
+&\frac{1}{2}\ket{4_H}\otimes(\frac{1}{\sqrt{2}}\ket{\uparrow_A\uparrow_B}+\frac{1}{\sqrt{2}}\ket{\downarrow_A\downarrow_B}).
\label{BHdStoy}
\end{eqnarray}
Here the space $H$ is the combined space of the black hole and de
Sitter stretched horizons. The states $\ket{n_H}$ are orthogonal
states in $H$. 
These four states do not span $H$ which has a very large dimension,
but maximal entanglement
between two systems can be expressed using only a number of states equal to the
dimension of the smaller of the two spaces. 

By inspection one can see that the state in Eqn.~\ref{BHdStoy} gives
a density matrix for the $A\otimes B$ space which has Bell states as
eigenstates. A
single line of this equation would be a state where A and B were
maximally entangled (as in
previous toy model), but overall there
exists no entanglement between A and B. One way to 
see this by is examining the correlations that would be observed between
$A$ and $B$. The density matrix of $ A \otimes B$ describes a uniform
statistical mixture of the states given in each line of the
equation. The first two lines show states where $A$ and $B$ are perfectly
anti-correlated and the last two are states where they are perfectly
correlated. Since half of the states are anti-correlated and half are
correlated, you expect no overall correlation between systems $A$ and $B$
and therefore no entanglement. 

Now consider an observer that has collected the radiation and wishes
to pass through the black hole's horizon. Just as before, to ensure
smooth spacetime we need a subsystem of $H$ to be entangled with system
$A$.  We label the subsystem of $H$ that participates in this
entanglement ``$C$''. A state for the whole system which gives such
entanglement is
\begin{eqnarray}
&\frac{1}{4}(\ket{\tilde{1}_H}\ket{c_1}+\ket{\tilde{2}_H}\ket{c_2})\otimes(\ket{\uparrow_A\downarrow_B}-\ket{\downarrow_A\uparrow_B})+ \nonumber\\
&\frac{1}{4}(\ket{\tilde{1}_H}\ket{c_3}+\ket{\tilde{2}_H}\ket{c_4})\otimes(\ket{\uparrow_A\downarrow_B}+\ket{\downarrow_A\uparrow_B})+ \nonumber\\
&\frac{1}{4}(\ket{\tilde{1}_H}\ket{c_5}+\ket{\tilde{2}_H}\ket{c_6})\otimes(\ket{\uparrow_A\uparrow_B}-\ket{\downarrow_A\downarrow_B})+ \nonumber\\
&\frac{1}{4}(\ket{\tilde{1}_H}\ket{c_7}+\ket{\tilde{2}_H}\ket{c_8})\otimes(\ket{\uparrow_A\uparrow_B}+\ket{\downarrow_A\downarrow_B}).
\end{eqnarray}
The $\ket{\tilde{n}_H}$'s are new orthogonal states in $H$ which now
only includes the de Sitter stretched horizon. The $\ket{c_m}$'s are linearly dependent states are in
$C$. Line by line, the apparent entanglement between $A$ and $B$ does
not seem to be disturbed (appearing just as in
Eqn.~\ref{BHdStoy}). This is as we expect since particles $A$ and $B$ 
have not interacted with anything. To see where the entanglement lies,
we need to consider the entire state. 

To get the required entanglement between $C$ and $A$ we choose
\begin{eqnarray}
&\ket{c_1}=-\frac{1}{\sqrt{2}}\ket{\uparrow_c}+\frac{1}{\sqrt{2}}\ket{\downarrow_c}, \ket{c_2}=-\frac{1}{\sqrt{2}}\ket{\uparrow_c}-\frac{1}{\sqrt{2}}\ket{\downarrow_c} \nonumber\\
&\ket{c_3}=\frac{1}{\sqrt{2}}\ket{\uparrow_c}+\frac{1}{\sqrt{2}}\ket{\downarrow_c}, \ket{c_4}=\frac{1}{\sqrt{2}}\ket{\uparrow_c}-\frac{1}{\sqrt{2}}\ket{\downarrow_c} \nonumber\\
&\ket{c_5}=-\frac{1}{\sqrt{2}}\ket{\uparrow_c}-\frac{1}{\sqrt{2}}\ket{\downarrow_c}, \ket{c_6}=\frac{1}{\sqrt{2}}\ket{\uparrow_c}+\frac{1}{\sqrt{2}}\ket{\downarrow_c} \nonumber\\
&\ket{c_7}=\frac{1}{\sqrt{2}}\ket{\uparrow_c}-\frac{1}{\sqrt{2}}\ket{\downarrow_c}, \ket{c_8}=-\frac{1}{\sqrt{2}}\ket{\uparrow_c}+\frac{1}{\sqrt{2}}\ket{\downarrow_c}.
\end{eqnarray}
Then the full state can be written as
\begin{eqnarray}
&(\frac{1}{\sqrt{2}}\ket{\uparrow_C\downarrow_A}-\frac{1}{\sqrt{2}}\ket{\downarrow_C\uparrow_A})\otimes \nonumber \\
&(\frac{1}{\sqrt{2}}(\ket{\uparrow_B}+\ket{\downarrow_B})\ket{\tilde{1}_H}+\frac{1}{\sqrt{2}}(\ket{\uparrow_B}-\ket{\downarrow_B})\ket{\tilde{2}_H}).
\end{eqnarray}
Now we can see explicitly that systems $C$ and $A$ are in a Bell state and are
therefore maximally entangled.  The entanglement with the horizons has conspired to protect the local physics
required to meet Postulate 4 in addition to the other three
postulates, thus evading the AMPS argument.

This basic argument holds at any point during the black hole's
evolution. The Page time, when the black hole has lost half its
entropy in our scenario hold no significance. The overall evolution of
the black hole is evolving from a mixed state to a mixed state. 

Where this differs from simply starting the black hole in some
entangled state (such as in ~\cite{Page:2013dx, Maldacena:2013xja}) is that by having the black hole be entangled with the
de Sitter horizon, we gain two useful features of that horizon;
its size and inaccessibility. The very high dimension of de Sitter
stretched horizon subspace means that
typical states for the whole system will have the other subsystems
highly entangled with the de Sitter horizon. While the states hidden
from outside observers in the black hole stretched horizon eventually
are revealed as the black hole decays, as long as the 
de Sitter space is stable we maintain a very large space of hidden
degrees of freedom throughout time in this scenario. The fact that no single
observer would encounter both the black hole stretched horizon and the
de
Sitter stretched horizon allows complementarity to never encounter
the AMPS contradictions. 

\section{The Role of Measurement}
\subsection{Background discussion}
A potentially confusing concept in this framework is the effect of
measurement in these thought experiments. We will mostly default to
the ``many worlds'' interpretation of quantum measurement (where the
only evolution of the wavefunction is the unitary evolution determined
by the Schroedinger equation, with no explicit ``collapse'') but much of the
discussion is not dependent on that choice. Understanding the role of
measurements in these thought experiments requires an understanding of
entanglements when measurements are made. One way to describe the
measurement is from the perspective of the observer. When a
good measurement is made the observer is only aware of a single
outcome (or ``Everett branch'') and doesn't see interference with
other possible outcomes that, from the perspective of the observer only ``could
have happened but didn't'' but which are still represented in the
complete wavefunction. Another description 
includes the observer as a part of the total wave function. In that
description, the observer becomes entangled with the system being
observed through the interactions that facilitate the measurement. It
is this entanglement (and the {\em stability} of this entanglement)  that
enables the first ``observer-centered'' description because the observer is correlated with
the measured system. Both descriptions will be important in what
follows. 

In the original firewall paper ~\cite{Almheiri:2012rt}, a thought experiment is
presented in which a measurement is being performed by an
observer.  In agreement with those authors ~\cite{Almheiri:2013hfa} we
believe that an actual observer making measurements is not needed 
for their argument to work. The important thing is to clearly identify
the frame from which the situation is being analyzed, and it is
convenient to identify frames by specifying observers. If the observer actually
does make certain measurements that can alter some details of the
discussion but not the main points. 
An observer that encounters emitted particles outside the black hole for long enough
and then falls in describes a frame in which the arguments given in
Sect.~\ref{sec:bh} produce a conflict between quantum mechanics and the
``no drama'' postulate.  If the no drama postulate holds, quantum monogamy would seem to be
contradicted since the early radiation should have parts that are
entangled with each other as well as the now revealed interior of the
black hole. This argument holds regardless of whether or not the
observer measures the radiation. In the case of no measurement,
entanglement of the black hole with the radiation presents the
contradiction. If the radiation is measured
then the problem results from entanglement of the black hole with the
measurement apparatus and observer. 

 \subsection{Quantum Teleporting a Firewall}
Other experiments can be performed that would rearrange
entanglements. One such experiment is a Bell measurement, which
projects arbitrary states onto a Bell state. This is the basis of
quantum teleportation. Consider two subsystems, one which an observer
interacts with (or ``controls'') and another with which the observer
does not interact at all. If the two subsystems are maximally
entangled, performing Bell measurements will create
entangled Bell states in the observed subsystem.  As a result the
observer will be certain that the non-observed subsystem will also be
in a particular Bell state related (via the properties of the initial
state) to the outcome of the measurement of the observed
subsystem.\footnote{Although ``Bell states'' are technically described
  as states in a four-state Hilbert space these points can be
  generalized to larger spaces.} Performing the proper coherent operation (or ``quantum 
computation'') on the controlled subsystem can ``teleport'' a
particular quantum state:  The state of the non-controlled system can
be modified without any direct interactions, only exploiting the
initial entanglement between the two systems~\cite{Bennett:1992tv}.

This concept can be applied to black holes as well. Consider in the
AMPS thought experiment, an observer that has collected enough
radiation such that she possesses a system that the black hole is
maximally entangled with. She could make measurements that would
entangle radiation particles with each other resulting in the black
hole's subsystems to become entangled with themselves. For particular
measurements, the entanglements desired for ``no drama'' can be
created and the black hole would be projected (or ``teleported'') into a state with no
firewall. This does not serve as a counter argument to firewalls. It
simply shows that through particular measurements a system can have
its state projected onto a particular subspace.
Much of the firewall discussion necessarily revolves
around which states one thinks are typical ~\cite{Marolf:2013dba,Almheiri:2013hfa,VanRaamsdonk:2013sza}, since there is general
agreement that states with and without firewalls exist. The AMPS paper stands by the notion that firewalls are
the typical state for black holes. But sufficiently elaborate
measurements to teleport to a non-firewall
state should be possible in principle. Likewise, even if you don't believe firewalls are
typical ~\cite{Nomura:2012sw,Nomura:2012ex,Nomura:2013gna,Harlow:2013tf} you could envision specific measurements that would project
the black hole into a firewall state.  Maldacena and Susskind have examined quantum computations
performed on a pair of entangled black holes, sending signals between
them, and have analyzed the features needed to send (or in our words ``teleport'') a firewall~\cite{Maldacena:2013xja,Susskind:2014rva}.

In our thought experiment in de Sitter space, the radiation emitted
from the black hole will have practically no entanglement with 
the black hole. The absence of this sort of entanglement eliminates the
possibility of teleportation and means that
measurements of the radiation after the black hole has 
formed and begun to decay will leave the black hole basically
unchanged. These measurements will 
affect the states of systems that {\em are} entangled with the
emitted radiation, that is, small subsystems of the de Sitter stretched horizon.

One can also consider measurements that can be made on the initial Gibbons-Hawking
radiation that came from the de Sitter horizon to form the black
hole. A careful measurement of this incoming radiation can result in a
pure state of radiation that will become a pure black hole (described
by states 
modeled by Eqn.~\ref{BHevapState}). This would be projecting
onto a state that will evolve into a firewall black hole state. This
is an example of using teleportation to project onto what we
consider in our picture to be an atypical firewall state. 

\section{Discussion and Conclusions}
\label{CaD}
The concept of black hole firewalls has arisen from the desire to
ensure completely unitary evolution, where pure states evolve into pure
states. Our toy model has not discarded unitary quantum mechanics, nor
have we abandoned the idea that the entire system is in a pure
state. But in our picture the black hole starts in a mixed state
due to entanglement with inaccessible states of the de Sitter space
stretched horizon. In this
case the information problem is nullified.
The initially completely mixed black hole would
contain no quantum information and as a result there is no information
for the radiation to carry away (the information would instead reside in the
entanglements, as discussed for example in~\cite{Albrecht:1994ay}) . The mixed nature of our states
removes all entanglement among subsystems of the black hole and its
decay products and as a result saves us
from worrying about quantum monogamy provided we cannot access the
system that these states are actually entangled with. While we have
used the stretched horizon of de Sitter as a simple illustration, our
conclusions should apply to any situation where a black hole is fully
entangled with a large inaccessible system.

We note that all the discussions of the firewall issue revolve around
the question of what is a typical state for a black hole.\footnote{Separately we have
	considered whether a very strong form of complementarity and
	holography could exist to enforce entanglements of horizons to
	absolutely forbid firewalls in all cases. But that would involve adding
	a new principle regarding non-local interactions between the stretched
	horizons of different objects, with some seemingly exotic consequences (such as black
	holes evolving from pure to mixed states, even as the whole
	system evolves unitarily).  Such a new principle is
	certainly not needed for the main points of this paper which are that
	with a large enough external space black hole firewalls can plausibly
	avoided based only on the statistical arguments about entanglement
	developed by Page.}
In an earlier preprint version of this paper we tried to argue that
the maximal entanglement we discuss here would be typical in certain
cosmological scenarios. We have since changed our minds about
this.  The fact that we, and the black holes we see around us are
all correlated with a cosmological state of the universe which is very
far from equilibrium prevents the maximal entanglement
we require for our analysis from being realistic in the universe we
observe.   The vast majority of
black holes formed in our universe would originate not from
fluctuations like we've considered in this paper, but instead from the
gravitational collapse of ``ordinary'' matter. The entropy of the
matter which will form a black hole of mass, $M$ will be on the order
of $M^{3/2}$ which is much smaller than entropy needed to start a
black hole in a maximally mixed state, $M^2$. For this reason, we
might expect black holes formed in many cosmological descriptions to
begin essentially pure, so much of the discussion in this paper will not apply. In
the language of this paper, even very basic information about the
universe (which for example supports a simple FRW description) constitutes
measurements which take us essentially all the way to the point of
``teleporting'' a firewall into the black holes around us, even if the
entire observed universe is entangled with something larger. An error
in our earlier thinking was to assume that the measurements required
for this teleporting would be extremely complicated ones of the sort
discussed in~\cite{Maldacena:2013xja,Susskind:2014rva}. Now we have
realized that such measurements have already been done.\footnote{We
  thank Don Marolf for pointing out this error to us.}
 
 Our approach may still be relevant in a dSE description of cosmology
 ~\cite{Albrecht:2004ke,Albrecht:2009vr,Albrecht:2011yg,Albrecht:2013lh}. If
 there was some strong form of holography in place, enforcing the
 Hilbert space of our entire universe to be small (dimension of order
 $e^{10^{12}}$) resulting in smaller Hilbert spaces actually needed to
 describe black holes, radiation and everything else we observe. However, such a
 concept has not been rigorously realized, and simple things we know
 about the properties (such as heat capacities and entropies) of
 everyday matter suggest that such a construction would be in conflict
 with known physics. 

We wrap up by considering briefly a  more ordinary implementation of our
ideas.  Consider a standard semiclassical picture of the
universe with quantum fields on a Friedmann-Robertson-Walker (FRW)
background (with no de Sitter stretched horizon, perhaps having an unstable
form of dark energy). If we consider the
state of the entire universe to be pure, a black hole of the sort we
actually observe will have a space of states that make up a small portion of the
total universe's very large Hilbert space. Using the Page arguments~\cite{Page:1993df} 
you'd expect subsystems that are less than half the space (or much
less, as is the case for a typical black hole in our universe) to
be essentially maximally mixed. At first glance it appears that this
mixing will provide all the same technical features we have used in this paper to
argue that firewalls are not likely, without evoking something as
exotic as a de Sitter stretched horizon.  

However, a key feature of
the de Sitter stretched horizon is its persistent inaccessibility to
the relevant observers. 
We do not expect there is any way for an
observer to find the information in the de Sitter
stretched horizon which is entangled with the black hole (a tiny fraction
of the bits in that stretched horizon) and then jump into the
hole. This feature allows us to evoke complementarity to avoid the
possibility of observing  quantum inconsistencies.
In the FRW case it is not at all clear we have this same feature.
Indeed, in a realistic universe there will be lots of ``inaccessible'' states in other
black hole horizons. These holes will eventually decay, but others
will form.  The question of
whether the stretched horizons of other black holes (or other aspects
of a realistic FRW picture) could play a role similar to the one we
have assigned to the de Sitter stretched horizon is an interesting one
which we do not address here. 

Our basic point is that if a black hole is entangled with a large
inaccessible space, the usual discussions of the firewall and
information problems are dramatically changed.  We have argued that in this case
there no longer is an information problem, and that one can satisfy
the four postulates of complementarity without any
contradictions. A de Sitter stretched horizon appears to have the
necessary features to play the role of this large space, and we have
used that for most of this paper to make our points. Our considerations may only
be applicable in the case of the largest space of possible black hole
initial states such as black holes forming from a fluctuation in de
Sitter space.  That would make our considerations inapplicable to
cosmologically realistic cases. However, we have also raised the
possibility that each black hole we observe could have sufficient
entanglement with other existing black holes to allow our arguments to
go through. If that were the case, observed black holes would not have
an information problem and the AMPS analysis would not apply.

\acknowledgements
We thank Don Page and Don Marolf for helpful discussions.  This work
was supported in part by DOE Grant DE-FG03-91ER40674.

\section*{Appendix}
\subsection{Entanglement and Quantum Monogamy}
In this section we review some basic properties of quantum pure and
mixed states and entanglement that we have utilized in this paper. A
pure quantum state is a state which can be represented by a single
vector in the Hilbert space (denoted by a ket). A mixed state is a
statistical mixtures of pure states and cannot be represented by a
single ket, instead it is described by a density matrix. The maximal
deviation from a pure state is called a maximally mixed state, which
appears as a diagonal density matrix with equal entries on the
diagonal in every basis. 

Quantum mixed states for
subsystems can occur due to entanglement. Page has argued that 
such entanglement is typical when
examining a subsystem of a larger system which is in a pure state. If
a system in a pure state is divided into two equal sized subsystems,
$A$ and $B$, that are both maximally mixed then systems $A$ and $B$ are
said to be maximally entangled with each other. The simplest examples
of maximal entanglement are the Bell states which span a four
dimensional Hilbert space, a product space of two two-state systems:
\begin{eqnarray}
&\frac{1}{\sqrt{2}}\ket{\uparrow_A\downarrow_B}-\frac{1}{\sqrt{2}}\ket{\downarrow_A\uparrow_B} \nonumber\\
&\frac{1}{\sqrt{2}}\ket{\uparrow_A\downarrow_B}+\frac{1}{\sqrt{2}}\ket{\downarrow_A\uparrow_B} \nonumber\\
&\frac{1}{\sqrt{2}}\ket{\uparrow_A\uparrow_B}-\frac{1}{\sqrt{2}}\ket{\downarrow_A\downarrow_B} \nonumber\\
&\frac{1}{\sqrt{2}}\ket{\uparrow_A\uparrow_B}+\frac{1}{\sqrt{2}}\ket{\downarrow_A\downarrow_B}.
\label{Bell}
\end{eqnarray}
Each state appears to be a pure state in $A \otimes B$ but for both $A$
and $B$ subsystems their states are maximally mixed. It is easy to see
the high level of correlations between systems $A$ and $B$ for these
states. For each of the above states taken individually, when spin is
measured in the basis shown, a measurement of spin for $A$
always correlates with a specific spin for $B$. These states represent the
strongest possible correlations between two two state systems which is why
they are described as being maximally entangled. 

One can consider a larger space by adding subsystem $C$ to the
picture.  One can construct product states between the Bell states for
$A \otimes B$ and pure states in $C$ without disturbing the
entanglement between $A$ and $B$, for example:
\begin{eqnarray}
(\frac{1}{\sqrt{2}}\ket{\uparrow_A\downarrow_B}-\frac{1}{\sqrt{2}}\ket{\downarrow_A\uparrow_B})\otimes\ket{\psi_C}.
\end{eqnarray}
System $A$ can also be described as being maximally entangled with
system $B \otimes C$ but not vice-versa. In general when a system is
maximally entangled with a larger system, a {\em subsystem} of the large
system can be identified which is maximally entangled with the small
system. For two subsystems to be mutually maximally entangled they need to be
the same size (as is the case with our illustration in Eqn.~\ref{Bell}).

Quantum monogamy can be stated as follows: If system $A$ is maximally
entangled with $B$ then $A$ cannot share any entanglement with another
system $C$. Assuming this is not true leads to a contradiction. We
will next demonstrate the weakest form of this statement (which we use in
this paper), namely that $A$ and $C$ cannot be maximally entangled.

By definition, $A$ and $B$ being maximally entangled means $A \otimes
B$ is in a pure state, and 
state of $A \otimes B$ can written as a single ket,
$\ket{\Psi_{AB}}$. Also, $A$ and $B$ will each be in maximally mixed
states. 

Suppose $A$ and $C$ were maximally entangled which would imply $A
\otimes C$ is in a pure state and $A$ and $C$ are both maximally
mixed. Then the state of $A \otimes C$ can written as a single ket,
$\ket{\Phi_{AC}}$. 

Combining the above two assumptions means the state for $A \otimes B
\otimes C$ can be written as $\ket{\Psi_{AB}} \otimes
\ket{\Phi_{AC}}$. Since this state can be written as a single ket, $A
\otimes B \otimes C$ is in a pure state.  
However, $A \otimes B$ being in a pure state while $C$ is in a mixed
state means $(A \otimes B) \otimes C$ is in a mixed state which is in
direct conflict with the previous statement. 
Simply put, the linearity of quantum mechanics prevents any state that
can be written down for $A \otimes B \otimes C$ that will give maximal
entanglement for $A$ and $B$ as well as the needed correlations
between $A$ and $C$.  

These properties of entanglement, combined with statistical arguments
about what is typical when a system is divided into subspaces lie at
root of the results developed by Page. The statistical arguments are
used to make the case that certain entanglements are likely to be
maximal, a feature we have simply put in by hand by using Bell states
in our illustration here.  A key result is that typically 
a pure state system divided into a large and small subsystem results in the
small subsystem being very mixed and entangled with the large
subsystem.  The large subsystem is more pure and most of its
entanglements are between its own subsystems and not with
the other small subsystem. 

\bibliography{AA}
\end{document}